\documentclass[8.5pt,twoside,twocolumn]{article}
\usepackage[super,sort&compress,comma]{natbib}
\usepackage{mhchem}
\usepackage{mathrsfs}
\usepackage{amsmath,amssymb,latexsym}
\usepackage{bm}
\usepackage{times,mathptmx}
\usepackage{sectsty}
\usepackage{balance}

\usepackage{graphicx} 
\usepackage{lastpage}
\usepackage[format=plain,justification=raggedright,singlelinecheck=false,font=small,labelfont=bf,labelsep=space]{caption}
\usepackage{fancyhdr}
\usepackage{color}
\begin{document}

\newcommand{\st}[1]{{\color{red} \bf{#1}}}
\newcommand{\lc}[1]{{\color{blue} \bf{#1}}}
\newcommand{\um}{~\mu\mathrm{m}}
\newcommand{\erf}{\mathrm{Erf}}
\newcommand{\re}{\mathrm{Re}}
\newcommand{\sinc}{\mathrm{sinc}}
\newcommand{\rect}{\mathrm{rect}}
\setcounter{secnumdepth}{5}

\thispagestyle{plain}
\fancypagestyle{plain}{
\fancyhead[L]{\includegraphics[height=8pt]{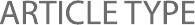}}
\fancyhead[C]{\hspace{-1cm}\includegraphics[height=20pt]{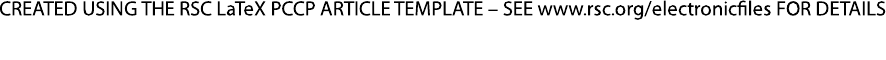}}
\fancyhead[R]{\includegraphics[height=10pt]{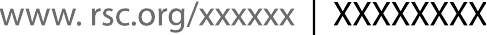}\vspace{-0.2cm}}
\renewcommand{\headrulewidth}{1pt}}
\renewcommand{\thefootnote}{\fnsymbol{footnote}}
\renewcommand\footnoterule{\vspace*{1pt}%
\hrule width 3.4in height 0.4pt \vspace*{5pt}}

\makeatletter
\def\subsubsection{\@startsection{subsubsection}{3}{10pt}{-1.25ex plus -1ex minus -.1ex}{0ex plus 0ex}{\normalsize\bf}}
\def\paragraph{\@startsection{paragraph}{4}{10pt}{-1.25ex plus -1ex minus -.1ex}{0ex plus 0ex}{\normalsize\textit}}
\renewcommand\@biblabel[1]{#1}
\renewcommand\@makefntext[1]%
{\noindent\makebox[0pt][r]{\@thefnmark\,}#1}
\makeatother
\renewcommand{\figurename}{\small{Fig.}~}
\sectionfont{\large}
\subsectionfont{\normalsize}

\fancyfoot{}
\fancyfoot[LO,RE]{\vspace{-7pt}\includegraphics[height=9pt]{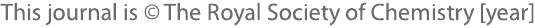}}
\fancyfoot[CO]{\vspace{-7.2pt}\hspace{12.2cm}\includegraphics{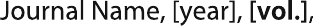}}
\fancyfoot[CE]{\vspace{-7.5pt}\hspace{-13.5cm}\includegraphics{RF}}
\fancyfoot[RO]{\footnotesize{\sffamily{1--\pageref{LastPage} ~\textbar  \hspace{2pt}\thepage}}}
\fancyfoot[LE]{\footnotesize{\sffamily{\thepage~\textbar\hspace{3.45cm} 1--\pageref{LastPage}}}}
\fancyhead{}
\renewcommand{\headrulewidth}{1pt}
\renewcommand{\footrulewidth}{1pt}
\setlength{\arrayrulewidth}{1pt}
\setlength{\columnsep}{6.5mm}
\setlength\bibsep{1pt}

\twocolumn[
  \begin{@twocolumnfalse}
\noindent\LARGE{\textbf{Probing shear-induced rearrangements in Fourier Space. II. Differential Dynamic Microscopy}}
\vspace{0.6cm}

\noindent\large{\textbf{S. Aime$^{\ast}$\textit{$^{a,b}$} and L. Cipelletti\textit{$^{a}$  }
}}\vspace{0.5cm}



\noindent \normalsize{We discuss in two companion papers how Fourier-space measurements may be coupled to rheological tests in order to elucidate the relationship between mechanical properties and microscopic dynamics in soft matter. In this second companion paper, we focus on Differential Dynamic Microscopy (DDM) under shear. We highlight the analogies and the differences with Dynamic Light Scattering coupled to rheology, providing a theoretical approach and practical guidelines to separate the contributions to DDM arising from the affine and the non-affine part of the microscopic displacement field. We show that in DDM under shear the coherence of the illuminating source plays a key role, determining the effective sample thickness that is probed. Our theoretical analysis is validated by experiments on 2D samples and 3D gels.}
\vspace{0.5cm}
\end{@twocolumnfalse}
]



\footnotetext{\textit{$^{a}$ L2C, Univ Montpellier, CNRS, Montpellier, France.\newline\textit{$^{b}$}Present address: School of Engineering and Applied Sciences, Department of Physics, Harvard University, Cambridge, USA.\newline E-mail: aime@seas.harvard.edu}}


\section{Introduction}

Soft matter is easily deformed by applying even modest loads. As we have argued in the first paper of this series~\cite{comp}, to which we refer for a more detailed bibliography survey, understanding the interplay between the structure, the microscopic dynamics and the rheological properties of soft materials is a field of great current interest, with implications both at a fundamental level and in many industrial applications~\cite{larson_structure_1998}. Experiments that probe \textit{simultaneously} the rheology and the structure or dynamics are particularly valuable, since the system response often exhibits complex spatio-temporal patterns that may vary significantly from run to run, especially in the nonlinear regime~\cite{divoux_transient_2010,bonn_yield_2017,knowlton_microscopic_2014,ghosh_direct_2017}.

While research in the past has been mostly devoted to the relationship between structure and mechanical properties, current efforts focus on the microscopic dynamics, which are often more sensitive than structural quantities to external perturbations~\cite{aime_microscopic_2018}. Direct-space measurements, e.g. by optical and confocal microscopy, have often been used~\cite{koumakis_yielding_2012,sentjabrskaja_creep_2015,derks_confocal_2004,besseling_three-dimensional_2007,schall_structural_2007,lin_multi-axis_2014}. As discussed in Ref.~\cite{comp}, scattering-based methods are an appealing alternative to direct-space techniques: while the former lack the ability to follow the trajectories of individual particles, they can be applied to a broader range of samples, allow a larger sample volume to be probed, and access a wider range of time and length scales.

Previous works based on scattering methods have usually adopted the so-called echo protocol, where the dynamics are probed stroboscopically at each cycle while the sample is submitted to an oscillating drive~\cite{hebraud_yielding_1997,petekidis_rearrangements_2002,rogers_echoes_2014,tamborini_plasticity_2014}. The echo protocol greatly simplifies the data analysis, since it directly yields the dynamics associated to irreversible rearrangements. In the more general case where the sample continuously deform during the experiment (e.g. in a creep test), one is confronted with the challenge of disentangling the trivial contribution to the dynamics due to the affine deformation field from the interesting part associated with non-affine motion and plasticity. This is of course an easy task in microscopy experiments, where the affine displacement can be subtracted off the particles' trajectories, but it requires special care in scattering experiments. In the first paper of this series~\cite{comp} we have shown how a Dynamic Light Scattering (DLS) setup may be coupled to a shear cell in order to probe affine displacements and non-affine rearrangements, providing guidelines for mitigating and correcting for the contributions to the DLS intensity correlation function arising from non-ideal conditions. Here, we present a similar study for Differential Dynamic Microscopy (DDM) under shear. DDM was first introduced in Ref.~\cite{cerbinoPRL}. It belongs to a family of techniques known as digital Fourier microscopy~\cite{giavazzi_digital_2014}, which combine features of both imaging and scattering: data are acquired in an image-forming setup (typically, a microscope), but they are processed in Fourier space. Similarly to DLS, the outcome of a DDM experiment are correlation functions describing the time and scattering-vector-dependent relaxation of density fluctuations.

DDM holds a great potential as a method to characterize dynamics under shear for several reasons: i) it is easily implemented in a commercial microscope; ii) like in conventional microscopy, subsets of the image can be independently analyzed, thereby providing one with a spatial map of the shear-induced dynamics; iii) unlike conventional microscopy, tracking the position of individual particles is not necessary (as a matter of fact, DDM does not even require individual particles to be resolved~\cite{buzzaccaro_ghost_2013}), which makes the method quite versatile; iv) the effective depth of focus can be easily tuned, which has important implications on measuring non-affine displacements, as we shall see.

Similarly to scattering techniques, DDM is a Fourier space method. One may thus argue that DDM correlation functions are very closely related to those measured by DLS, such that the methods presented in Ref.~\cite{comp} and the conclusions reached therein still hold. While this is broadly speaking true, there are several important distinctions that have to be taken into account to quantitatively exploit DDM under shear. These distinctions stem from two key differences with respect to DLS: first, in its original and most common formulation DDM is a heterodyne method, where both the transmitted illuminating beam and the light scattered by the sample are collected by the microscope objective. Second, in DDM the sample is typically illuminated by white light, with a coherence length generally smaller than the sample thickness, rather than by highly coherent laser light as in DLS. The goal of this work is to explore the consequences of these peculiarities, providing the reader with guidelines to correctly analyze DDM data taken under shear.

The rest of the paper is organized as follows: in Sec.~\ref{sec:MaterialsMethods} we briefly introduce the samples used for the measurements and the setup, and define the correlation functions used in DDM. In Sec.~\ref{sec:DDMresults} we develop a theoretical model of DDM under shear, starting from a discussion of the effect of a simple translation of the sample, and then presenting results for a purely affine deformation and the general case of both affine and non-affine microscopic dynamics. At each step, we show experimental data that validate our theoretical approach. We recapitulate our main findings and make some concluding remarks in Sec.~\ref{sec:conclusions}. A list of the symbols used in this paper is given for convenience in Appendix 1. Appendix 2 presents a detailed derivation of the correlation functions introduced in Sec.~\ref{sec:MaterialsMethods} and briefly discusses the impact of experimental noise on their normalization.

\section{Materials and methods}
\label{sec:MaterialsMethods}

\subsection{Samples}
The same samples introduced in Ref.~\cite{comp} were also used for the DDM measurements. They include frosted glass slides (2D samples) and polyacrylamide (PA) gels (3D samples), to which $\mathrm{TiO}_2$ nanoparticles of diameter $0.5~\um$ were added, at a volume fraction of about $0.01\%$. The typical thickness of the PA gels is $500\um$.

\subsection{Experimental apparatus}

The experiments were performed on a Leica DM IRB inverted microscope equipped with a 10x objective and a CCD camera (model DMK 23U274, from The Imaging Source GmbH). The CCD sensor has 1024x1024 pixels; the pixel size is $l_p=4.4\um$, corresponding to $l_p/M=0.44\um$ in the sample plane, $M=10$ being the microscope magnification. Microscopy under shear was performed by coupling the microscope to a custom-made stress-controlled shear cell~\cite{aime_stress-controlled_2016}, working in the sliding parallel plate geometry.

\subsection{Differential Dynamic Microscopy}
\label{sec:DDM}
For a thorough description of DDM, we address the reader to Refs.~\cite{cerbinoPRL,Giavazzi2009,giavazzi_digital_2014}.
Here, we simply recall that in DDM a time series of images taken with a microscope are processed in Fourier space, as explained succinctly below. The method works under different kinds of illumination; in this work we use Koehler illumination~\cite{giavazzi_digital_2014}. Throughout this paper we will use a Cartesian reference frame where the shear, vorticity, and shear gradient directions are $\hat{u}_x$, $\hat{u}_y$, and $\hat{u}_z$, respectively. $\hat{u}_z$ also corresponds to the optical axis direction, with $z=0$ the position of the object plane.
Under these conditions and for typical colloidal objects, the  microscope images result from the interference between the transmitted light and light scattered by the sample. The images contain $N_x \times N_y$ pixels, which record the intensity $I(\bm{x'})=\left|\bm{E}(\bm{x'})\right|^2$, where $\bm{x'}=(x', y')$ is the coordinate in the sensor (image) plane, $\bm{E}$ the total electric field, and where we have disregarded inessential prefactors. The total electric field $\bm{E}$ can be expressed as a sum of the transmitted beam $\bm{E}_{LO}$ and the scattered electric field $\bm{E}_{sc}$.

In the imaging geometry, $\bm{x'}$ is conjugated with a point in the sample (object) plane $\bm{x}$, which has coordinates $(x, y) = (x'/M, y'/M)$. In this geometry and assuming a weakly scattering sample (first Born approximation~\cite{BernePecora}), $\bm{E}_{LO}(\bm{x'})$ corresponds to the incident beam in $\bm{x}$ (local oscillator), whereas $\bm{E}_{sc}(\bm{x'})$ results from the interference of the light scattered by a set of particles belonging to an effective scattering volume centered around the object plane and sketched in Fig.~\ref{fig:focal_depth}f, which we shall discuss in depth in the following. The electric field in the sensor plane may then be expressed as
\begin{equation}
\bm{E}(\bm{x'})=\bm{E}_{LO}(\bm{x})+\sum\limits_i \bm{E}_{LO}(\bm{x}_i)K_{z_i}(\bm{x}-\bm{x}_i)\,,
\label{eqn:field_sum}
\end{equation}
where the sum runs on all particles in the sample, $(\bm{x}_i, z_i)$ is the 3D coordinate of $i$-th particle, and $\bm{E}_{LO}(\bm{x}_i)$ is the electric field incident on the $i$-th particle. The kernel $K_{z_i}(\bm{x}-\bm{x}_i)$ vanishes for particles outside the effective scattering volume; it depends on the scattering properties of the particle (the scattering form factor~\cite{BernePecora}), the optical path of the scattered light and the properties of the imaging system, through the microscope transfer function.

It is instructive to consider the expression of the kernel $K$ for the simple case of point-like particles illuminated by uniform, coherent light with wavevector $\bm{k}_{in}$, further assuming a uniform microscope transfer function. Under these assumptions, one finds $K_{z_i}(\bm{r}_i)=Se^{i\phi(\bm{r}_i,z_i)}$, with $S\ll 1$ a non-dimensional scattering cross-section, $\bm{r}_i=\bm{x}-\bm{x}_i$, and $\phi(\bm{r}_i,z_i)$ the phase difference (evaluated in $\bm{x}$) between the illuminating field and the light scattered by the $i$-th particle. By defining the wavevector $\bm{k}_{sc, i}=\bm{r_i}|\bm{k}_{in}|/|\bm{r_i}|$ of the scattered light and using simple geometric arguments, one finds that $\phi(\bm{r}_i, z_i)=(\bm{r}_i+z_i \hat{u}_z)\cdot(\bm{k}_{sc, i}-\bm{k}_{in})$, which draws the analogy between Eq.~\ref{eqn:field_sum} and the electric field in a heterodyne scattering experiment, with scattering vector $\bm{q}=\bm{k}_{sc, i}-\bm{k}_{in}$.
Note in particular that, as a consequence of coherent illumination, in this case $K_{z_i}(\bm{r}_i)$ does not contain any explicit dependence on the position $z_i$ of the particle along the optical axis, except for the propagation term in the phase factor $\phi(\bm{r}_i, z_i)$. This is not the case in a typical microscopy experiment, where only the signal from particles close to the object plane ($z=0$) is properly collected by the sensor.
In this case $K_{z_i}(\bm{r}_i)$ assumes a more complex structure, which is more easily expressed in Fourier space, where a finite focal depth $L_f(\bm{q})$ can be defined, as we will see in the following.
	
By introducing the local particle density operator $n(\bm{x}, z)=\sum_i\delta_D(\bm{x}-\bm{x}_i, z-z_i)$ with $\delta_D(\bm{x}, z)$ Dirac's delta function, one can cast Eq.~\ref{eqn:field_sum} into an integral that takes a particularly simple form under the assumption of a quasi-uniform illuminating field:
\begin{equation}
\bm{E}(\bm{x'})=\bm{E}_{LO}(\bm{x})\left[1+\int dz\int d_2r K_z(\bm{x}-\bm{r})n(\bm{r}, z)\right]\,.
\label{eqn:field_integral}
\end{equation}
Equation \ref{eqn:field_integral} can be used to calculate the intensity $I(\bm{x'})=\left|\bm{E}(\bm{x'})\right|^2$:
\begin{equation}
I(\bm{x'}) = \bm{I}_{LO}(\bm{x})\left[1+ \int dz \int d_2r 2\Re \left[K_z(\bm{x}-\bm{r})\right]n(\bm{r}, z)\right]\,,
\label{eqn:intensity}
\end{equation}
where $\bm{I}_{LO}(\bm{x})=\left|\bm{E}_{LO}(\bm{x})\right|^2$, $\Re$ denotes the real part, and where the quadratic term in the scattered field has been disregarded since in the weakly scattering regime $|\bm{E}_{sc}| << |\bm{E_{LO}}|$. Equation \ref{eqn:intensity} links $K_z(\bm{x})$ to the so-called point spread function $\textrm{PSF}_z(\bm{x})$ \footnote{In optical microscopy, it is customary to define the PSF as the setup intensity response to a point-like source. Thus, the usual PSF is proportional to the term $2\Re \left[K_z(\bm{r})\right]$ introduced in Eq.~\ref{eqn:intensity}.}, describing the response of the imaging system to a point source or point object placed at distance $z=\bm{x}\cdot\hat{u}_z$ from the object plane~\cite{giavazzi_digital_2014}.

Note that the convolution integral of Eq.~\ref{eqn:intensity} sets a linear relation between the intensity $I(\bm{x'})$ and the local concentration $n(\bm{r})$, which is precisely the kind of signal-concentration relation required for DDM to be applicable~\cite{giavazzi_digital_2014}. The first step of the DDM analysis consists in calculating the spatial Fourier transform\footnote{Here and in the following, we denote by $\tilde{f}(q_x,q_y)$ the 2D spatial Fourier transform of the function $f(\bf{x'})$.} of the intensity, $\tilde{I}(\bm{q})=\sum_p I(\bm{x'}_p)e^{-i\bm{q}\cdot\bm{x'}_p}$, where the sum runs on all sensor pixels, $\bm{x'}_p$ is the position of the $p$-th pixel and $\bm{q}$ represents a scattering vector, with $x, y$ components equal to $q_{x,y}=2\pi n_{x, y}M/(N_{x, y}l_p)$, with $n_{x, y}$ a couple of integer numbers. Note that a factor $M$ has been introduced in the definition of $\bm{q}$, in order to express the scattering vector in the units associated to the object plane, rather than the magnified units of the image plane. The Fourier transform $\tilde{I}(\bm{q})$ is directly related to the scattered intensity usually measured in a DLS experiment (see Appendix 2 for a more rigorous discussion). The temporal evolution of $\tilde{I}$ is quantified by correlating the Fourier transforms of pairs of images taken at time $t$ and $t+\tau$, respectively. Below, we define various degrees of correlations that are used in the data analysis.

We first introduce the `standard' degree of correlation used in DDM~\cite{cerbinoPRL,giavazzi_digital_2014}:
\begin{equation}
c_{DDM}(\bm{q},t,\tau)=1-\frac{\left<\left|\tilde{I}(\bm{q},t)-\tilde{I}(\bm{q},t+\tau)\right|^2\right>_{\bm{q}}}{\left<\left|\tilde{I}(\bm{q},t)\right|^2\right>_{\bm{q}}+\left<\left|\tilde{I}(\bm{q},t+\tau)\right|^2\right>_{\bm{q}}}\,,
\label{eqn:c_DDM}
\end{equation}
where the numerator of the second term in the r.h.s. is the image structure function originally introduced in Ref.~\cite{cerbinoPRL}. Here, $<\cdot \cdot \cdot>_{\bm{q}}$ denotes an average over a small region in Fourier space, centered around the wave vector $\bm{q}$, which will be chosen to be either parallel ($\bm{q}_x$) or perpendicular ($\bm{q}_y$) to the shear direction.
Provided that the scattered field $\tilde{\bm{E}}_{sc}(\bm{q}, t)$ dominates over the static background $\tilde{\bm{E}}_{LO}(\bm{q})$
\footnote{For the realistic case of a nearly uniform illumination, $\tilde{\bm{E}}_{LO}(\bm{q}) \approx 0$, except in the $q \rightarrow 0$ limit.}, it is shown in Appendix 2 that Eq.~\ref{eqn:c_DDM} reduces to the intermediate scattering function, or field correlation function~\cite{BernePecora}:
\begin{equation}
c_{DDM}(\bm{q},t,\tau)\approx \frac{\left<\tilde{\bm{E}}_{sc}(\bm{q}, t)\tilde{\bm{E}}_{sc}^*(\bm{q}, t+\tau)\right>_{\bm{q}}}{\left<\left|\tilde{\bm{E}}_{sc}(\bm{q},t)\right|^2\right>_{\bm{q}}} \,.
\label{eqn:intermediate_scattering}
\end{equation}
If one further assumes a stationary process, the statistics may be improved by averaging over $t$, yielding the DDM intensity correlation function
\begin{equation}
g_{2,DDM}(\bm{q},\tau)-1 = \left | \left <c_{DDM}(\bm{q},t+\tau)\right>_t \right |^2\,.
\label{eqn:g2_DDM}
\end{equation}

It is convenient to introduce also modified versions of the above correlation functions that correspond to the far-field correlation functions measured in DLS. These far-field DDM (FF-DDM) functions were already introduced in Ref.~\cite{philippe_efficient_2016}: as we shall show it, they are essential in order to efficiently analyze DDM data under shear. These functions are
\begin{equation}
\begin{split}
c_{FF-DDM}(\bm{q},t,\tau)&=\frac{\left<\left|\tilde{I}(\bm{q},t)\tilde{I}(\bm{q},t+\tau)\right|^2\right>_{\bm{q}}}{\left<\left|\tilde{I}(\bm{q},t)\right|^2\right>_{\bm{q}}\left<\left|\tilde{I}(\bm{q},t+\tau)\right|^2\right>_{\bm{q}}}-1 \\
	&=\frac{\left<\tilde{I}_{sc}(\bm{q}, t)\tilde{I}_{sc}(\bm{q}, t+\tau)\right>}{\left<\tilde{I}_{sc}(\bm{q},t)\right>^2_{\bm{q}}}
\label{eqn:c_FF-DDM}
\end{split}
\end{equation}
\begin{equation}
g_{2,FF-DDM}(\bm{q},\tau)-1 = \left <c_{FF-DDM}(\bm{q},t+\tau)\right>_t
\label{eqn:g2_FF-DDM}
\end{equation}
where $\tilde{I}_{sc}=|\tilde{\bm{E}}_{sc}|^2$, and the second line of Eq. \ref{eqn:c_FF-DDM} is again derived in Appendix 2 under the same conditions as for Eq.~\ref{eqn:intermediate_scattering}, i.e. that the scattered field dominates over the Fourier transform of the static background.
All the quantities introduced in Eqs.~\ref{eqn:c_DDM}-\ref{eqn:g2_FF-DDM} decay in principle from 1 at $\tau = 0$ to 0 for fully uncorrelated images. As a consequence of electronic camera noise and any deviation from a perfectly uniform illumination, however, experimentally measured correlation functions actually decay from a short-time value $1-\varepsilon_0(\bm{q})$ for $\tau\rightarrow0^+$ to a baseline $\varepsilon_\infty(\bm{q}) \gtrsim 0$. As explained in Appendix 2, a simple vertical shift and rescaling can correct for those effects as long as $\varepsilon_0(\bm{q})<<1$ and $\varepsilon_\infty(\bm{q})<<1$, which is usually the case. Such correction has been applied systematically to the experimental data shown in the following.

\section{Differential Dynamic Microscopy for a sheared sample}
\label{sec:DDMresults}

In analogy to Ref.~\cite{comp}, we model a 3D sample by a set of $\Sigma_z$ planes perpendicular to the shear gradient direction $\hat{u}_z$ and indexed by their distance $z$ from the object plane, for which $z=0$. For a sample undergoing a purely affine shear deformation with strain $\gamma$, each $\Sigma_z$ plane rigidly translates by $\bm{\delta}(z) =\gamma (z-z_s) \hat{u}_x$, $z_s$ being the $z$ coordinate of the stagnation plane of the shear deformation. In this section, we will first discuss the DDM signal from the rigid translation of an individual $\Sigma_z$ plane. Then, we will combine the signal of different $\Sigma_z$ planes to obtain the DDM correlation function for a purely affine shear deformation of the sample. Finally, we will discuss the contributions arising from any additional, non-affine displacement.

\subsection{Rigid translation of a 2D sample}
\label{subsec:DDMtrasl}

The DDM signal associated to a rigid translation $\bm{\delta}$ of a 2D sample can be computed by separating the electric field $\bm{E}(\bm{x'}, \bm{\delta})$ on the sensor in the sum of the (static) transmitted field $\bm{E}_{LO}(\bm{x'})$ and the $\bm{\delta}$-dependent contribution $\bm{E}_{sc}(\bm{x'}, \bm{\delta})=\bm{E}_{sc}(\bm{x'}-\bm{\delta'}, 0)$, which will translate rigidly by $\bm{\delta'}=M\bm{\delta}$ when the sample is translated by $\bm{\delta}$, $M$ being the system magnification.
Fourier transforming a shifted function introduces a phase term whose real part yields a cosine factor (see e.g. Ref.~\cite{GoodmanIntroductionFourierOptics2005}); one then finds for the conventional DDM intensity correlation function

\begin{equation}
    g_{2, DDM}(\bm{q}, \bm{\delta})-1 = \cos^2 \left(\bm{q} \cdot \bm{\delta}\right) \,,
    \label{eqn:DFMtrasl}
\end{equation}
where we have neglected the effect of pixel discretization and the finite size of the field of view, implicitly assuming $|\bm{\delta}| << L$, with $L=N_xl_p/M$ the size in the $\hat{u}_x$ direction of the sample portion that is imaged.

\begin{figure}[htbp]
\centering
  \includegraphics[width=\columnwidth,clip]{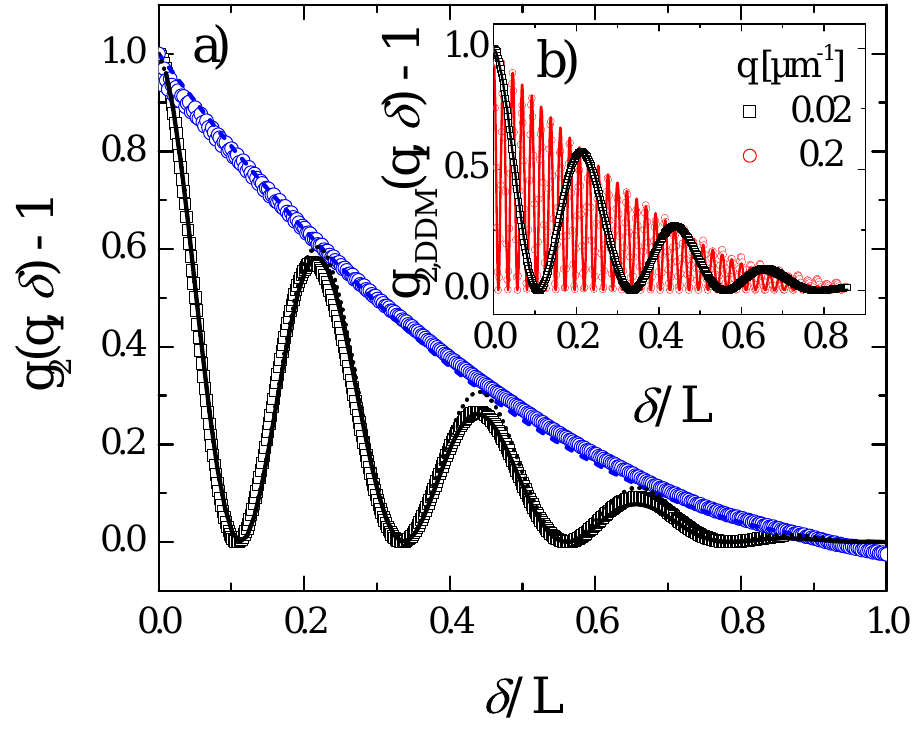}
  \caption{a) Comparison between $g_{2, DDM}-1$ (black squares) and $g_{2, FF-DDM}-1$ (blue circles) measured on a rigidly translating sample for the same scattering vector $q=0.02\um^{-1}$. Symbols are experimental data, lines are theoretical expectations based on Eqs. \ref{eqn:DFMtraslCorr} (black dotted line), \ref{eqn:DFMtraslCorr2} (black solid line) and Eq. \ref{eqn:DFMtraslCorr_pro} (blue dashed line). b) Symbols: $g_{2, DDM}-1$ evaluated for two different scattering vectors, as shown by the legend). Lines: theory based on Eq. \ref{eqn:DFMtraslCorr}.}
  \label{fig:trasl_dfm_pro}
\end{figure}

Equation~\ref{eqn:DFMtrasl} sets a first sharp difference between DDM and DLS: in Ref.~\cite{comp} we showed that in DLS any sample translation can be neglected as long as it is small enough with respect to the size of the illuminated sample. By contrast, Eq.~\ref{eqn:DFMtrasl} indicates that in the same regime the DDM signal is strongly oscillating. Such oscillations are clearly visible in experiments where $g_{2, DDM}-1$ was measured for a translating frosted microscope slide, see the red and black symbols in Fig.~\ref{fig:trasl_dfm_pro}b. Another remarkable feature seen in Fig.~\ref{fig:trasl_dfm_pro}b is the damping of the oscillations: this is due to the finite size effects disregarded in Eq.~\ref{eqn:DFMtrasl}. As $\delta$ grows, the overlap between the pair of images that are correlated decreases, eventually vanishing. Assuming that the field of view is homogeneously illuminated, the overlapping portion is perfectly correlated, while the non-overlapping one is fully uncorrelated. Accordingly, finite size effects result in a reduction of $c_{DDM}$ proportional to the overlap between the two images that are correlated, yielding:

\begin{equation}
    g_{2, DDM}(\bm{q}, \bm{\delta})-1 = \cos^2\left(\bm{q} \cdot \bm{\delta} \right)\left(1- \frac{\left|\delta\right|}{L}\right)^2
    \label{eqn:DFMtraslCorr}
\end{equation}

The black dotted lines in Fig~\ref{fig:trasl_dfm_pro}a show $g_{2,DDM}-1$ calculated from $c_{DDM}$ as predicted by Eq.~\ref{eqn:DFMtraslCorr}: a good agreement with the data is found, with no fitting parameters. Note however that the  measured oscillations appear to be slightly overdamped with respect to the prediction of Eq.~\ref{eqn:DFMtraslCorr}. This effect can be understood by taking into account the finite $q_x$ range over which the correlation function has been averaged, see Eq.~\ref{eqn:c_DDM}. 
Such effect can be taken into account by averaging Eq.~\ref{eqn:c_DDM} on a suitable interval $[q_x-q_w/2, q_x+q_w/2]$, which yields:

\begin{equation}
    g_{2, DDM}(\bm{q}, \bm{\delta})-1 = \cos^2\left(\bm{q} \cdot \bm{\delta} \right)\sinc^2\left(\frac{q_w\delta}{2}\right)\left(1- \frac{\left|\delta\right|}{L}\right)^2
    \label{eqn:DFMtraslCorr2}
\end{equation}

Note that $q_w$ cannot be vanishingly small, since the finite value of $L$ is associated to a Fourier transform discretized in units of $q_{min}=2\pi M/L$, which plays the role of the speckle size discussed in Ref. \cite{comp}.
Equation~\ref{eqn:DFMtraslCorr2} with $q_w=q_{min}$ is represented as solid lines in Fig~\ref{fig:trasl_dfm_pro}, showing an excellent agreement with the experimental data.

The oscillations resulting from the $\cos^2\left(\bm{q} \cdot \bm{\delta} \right)$ term undoubtedly complicate the analysis of the DDM data, especially since they depend on both the shift and the scattering vector. It is therefore convenient to consider the alternative far field correlation functions, Eqs.~\ref{eqn:c_FF-DDM} and \ref{eqn:g2_FF-DDM}. These correlation functions involve the modulus of the Fourier transform of the intensity: they are therefore insensitive to the phase factor introduced in Fourier space by a translation in real space. One simply finds

\begin{equation}
    g_{2, FF-DDM}(\bm{q}, \bm{\delta}) - 1= \left(1 -  \frac{\left|\delta\right|}{L} \right)^2
    \label{eqn:DFMtraslCorr_pro}
\end{equation}
which is the finite size correction factor introduced by the last term of Eq.~\ref{eqn:DFMtraslCorr}. We thus  obtain a $q$-independent correlation function, similarly to Eq.~9 in Ref.~\cite{comp}. The different functional form in DDM as compared to DLS is simply due to the fact that here the intensity profile of the illuminating field is assumed to be flat, instead of Gaussian as in DLS. The blue circles in Fig.~\ref{fig:trasl_dfm_pro}a show $g_{2, FF-DDM}-1$ for the same experiment with a translating frosted glass discussed previously. The far field correlation function exhibits no oscillations and is very well reproduced by the theoretical expression, Eq.~\ref{eqn:DFMtraslCorr_pro} (dashed line), without fitting parameters.

To summarize this section, we have shown that a rigid translation of a 2D sample results in inconvenient $q$-dependent oscillations of conventional DDM correlators. Using the far-field DDM correlation functions removes this difficulty, yielding results similar to those for DLS.

\subsection{Affine deformation of a 3D sample}
\label{affine_deformation}

In the linear regime, a homogeneous, solid-like sample sheared by $\gamma$ is deformed in a purely affine way: the displacement of a particle at position $(\bm{r}, z)$ is then

\begin{equation}
    \Delta\bm{r}(\bm{r}, z;\gamma)=\Delta\bm{r}^{\mathrm{(aff)}}(\bm{r}, z; \gamma) = (z - z_s) \gamma \hat{u}_x\,.
    \label{eqn:AffineDeformation}
\end{equation}

Note that all particles with the same $z$ coordinate (i.e. belonging to a given $\Sigma_z$ plane) translate rigidly by the same amount $\bm{\delta}(z) = \gamma (z - z_s) \hat{u}_x$. As a consequence, the DDM signal from an affine deformation contains a term coming from the overall average rigid translation of the sample, modulated by an additional term accounting for the relative motion of the different $\Sigma_z$ planes.

If the depth of focus is much larger than the gap $e$, one recovers the heterodyne scattering result, for which

\begin{equation}
    g_{2,DDM}(\bm{q}, \gamma) - 1 = \cos^2(q_x \gamma z_s)\sinc^2\left(q_x \gamma \frac{e}{2}\right)\,,
    \label{eqn:DFMshear_naiveCorr}
\end{equation}
where finite size corrections were omitted for clarity.

The $\cos^2(q_x \gamma z_s)$ term is analogous to the r.h.s. of Eq.~\ref{eqn:DFMtrasl}: it represents the contribution to the decay of the correlation function due to the average displacement in the shear direction. This contribution drops out for the fully symmetric case $z_s=0$, where the average displacement vanishes because the upper and lower plate of the shear cell move in opposite directions by the same amount. However, in the general case $z_s \neq 0$, the $\cos^2$ term introduces oscillations that complicate the data analysis, similarly to the case of Fig.~\ref{fig:trasl_dfm_pro}. Therefore, in the following we will exploit the advantages of the \textit{far field} correlation functions.
In principle, this approach yields the same results as the ones discussed in Ref. \cite{comp}. However, a key difference is that in microscopy, due to the finite depth of focus of the imaging system, only planes within a finite distance $L_f$ from the object plane will significantly contribute to the detected signal. This is a consequence of the limited coherence of the illuminating beam in DDM~\footnote{DDM typically uses white light, which for an extended source has sub-micron coherence length. Koehler illumination and a condenser (aperture) diaphragm are used to reduce the size of the illuminating source~\cite{Giavazzi2009}, thereby enhancing coherence, such that $L_f \approx 10-1000\um$.}, which has to be contrasted with the highly coherent laser light used in DLS.

Intuitively, particles close to the object plane will be sharply imaged, whereas their image will gradually blur as their distance to the object plane increases. In Fourier space, this corresponds to a change of the power spectrum that first impacts the high spatial frequencies (large $q$) and then gradually extends to include lower frequencies (small $q$). In other words, the relative contribution to the total signal coming from a scatterer at a given distance $z$ from the object plane varies according to the $q$ vector that is analyzed. Conversely, a given $q$ component of Fourier-transformed microscopy images consists mainly of the contribution of particles that are closer to the objet plane than a $q$-dependent depth of focus $L_f(q)$.

\begin{figure}[htbp]
\centering
  \includegraphics[width=\columnwidth,clip]{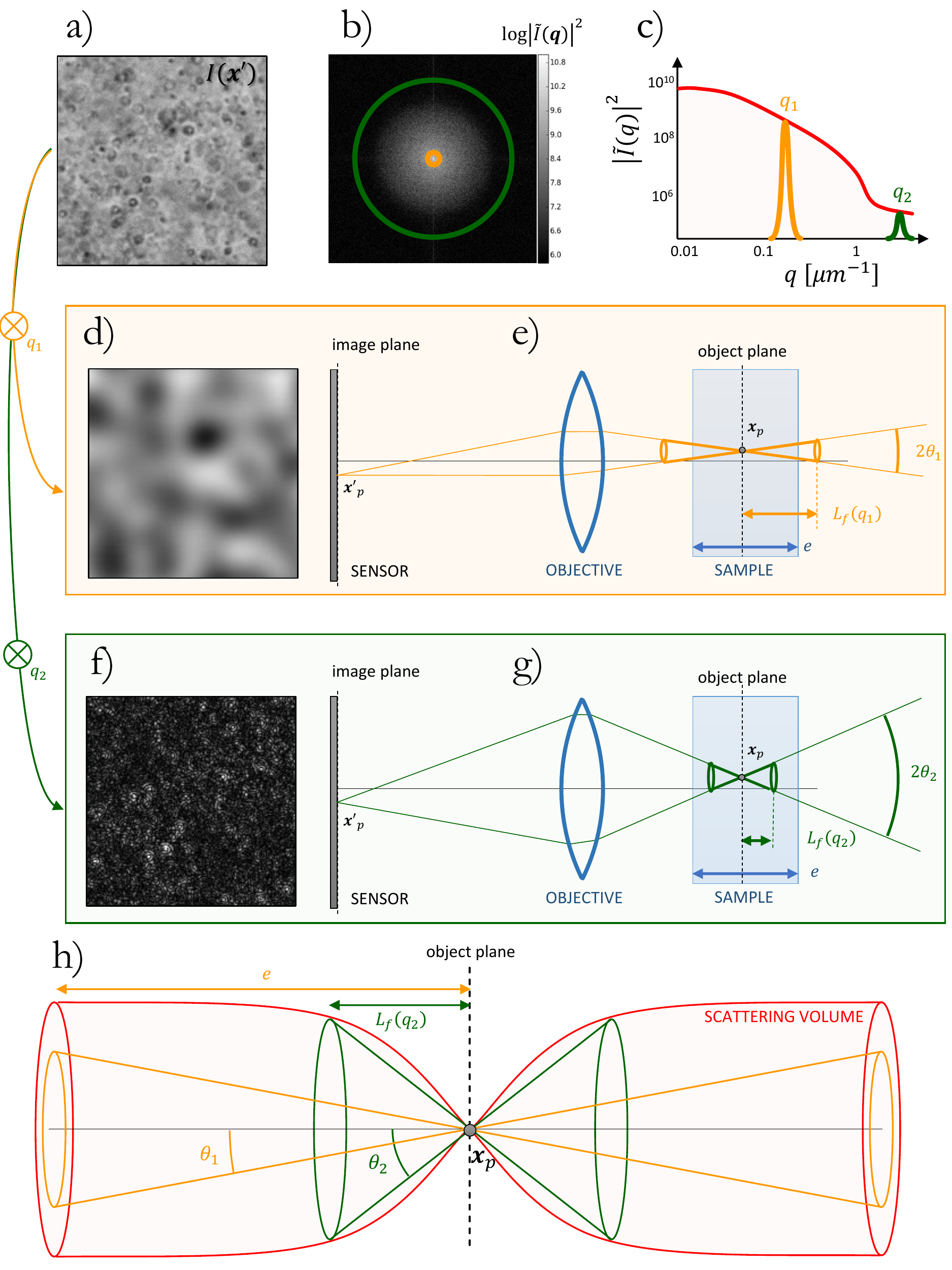}
  \caption{a) Representative sample image $I(\bm{x'})$. b) 2D Fourier power spectrum $\left|\tilde{I}(\bm{q})\right|^2$ of the image shown in a). The orange and green circles are representative scattering vectors with magnitude $q_1 = 0.2\um^{-1}$ and $q_2  = 4\um^{-1}$, respectively. c) Red line: azimuthally-averaged power spectrum $\left|\tilde{I}(q)\right|^2$. The orange and green peaks indicate Fourier bandpass filters extracting the contribution of scattering vectors centered about $q_1$ and $q_2$. d) Same image as in a), but filtered in Fourier space so as to retain only the contribution of scattering vectors close to $q_1$. e) Optical layout of the rays contributing to the filtered image shown in d), which form an angle $\theta_1 = 0.8^\circ$ with the optical axis. $L_f(q_1)$ is the depth of focus discussed in the text: note that at small $q$ the in-focus region encompasses the whole sample. f) and g): same as d) and e), but for scattering vector $q_2 > q_1$. Note that the in-focus region is smaller than the sample thickness.  h) Effective scattering volume obtained by integrating the conical surfaces associated to all scattering vectors accepted by the objective. All particles within the effective scattering volume contribute to the overall signal observed in position $\bm{x'}_p$ of image a).}
  \label{fig:focal_depth}
\end{figure}

This is sketched in Fig.~\ref{fig:focal_depth}. Panel a) shows a detail of one representative sample image, with grayscale encoding the intensity profile $I(\bm{x'})$. Particles have a diameter of $0.5\um$, thus they are large enough to be individually resolved. Different sizes of the particles imaged are related to different $z$ positions of the particles: the sharper the particle, the closer it is to the object plane. Such effect is reflected on the 2D Fourier power spectrum $\left|\tilde{I}(\bm{q})\right|^2$ (panel b, $\bm{q}=0$ at the center): particles of that size are expected to scatter light in a nearly $q$-independent fashion on the probed range of wavevectors, whereas the actual power spectrum drops by several orders of magnitude as $q$ is increased. This is better observed in panel c), where the power spectrum was averaged over the azimuthal angle. Here one can clearly see that smaller $q$ vectors contribute to the total signal much more than larger ones. The reason is two-fold: first, the intensity scattered by colloidal objects quite generally decreases with $q$. Second, the effective size of the scattering volume decreases with $q$, because the depth of focus $L_f(q)$ is smaller for Fourier components at larger $q$. To illustrate this, we compare the Fourier components corresponding to two different scattering vectors $q_1=0.2\um^{-1}$ and $q_2=4\um^{-1}$ in Fig.~\ref{fig:focal_depth}d-g. Panels d) and f) show the original image filtered in Fourier space, in order to isolate the contribution of scattering vectors $\left|\bm{q}\right|=q_1$ (top) and $\left|\bm{q}\right|=q_2$ (bottom), or, equivalently, the intensity scattered at angles $\theta_1$ and $\theta_2$, respectively (see panels e) and g)). As a consequence, the intensity recorded by the pixel $p$ in position $\bm{x_p'}$ in each filtered image originates from a very well defined region of the sample, sketched in panel h): a conical surface centered in the point $\bm{x_p}$ of the object plane conjugated with $\bm{x_p'}$ and with aperture $2\theta$. In this representation, the $q$-dependent depth of focus $L_f(q)$ plays the role of the height of the cone: the high-frequency image ($\left|\bm{q}\right|=q_2$, panel g)) only selects particles close to the object plane, implying a depth of focus $L_f(q_2)$ much smaller than the sample thickness $e$, whereas density fluctuations coming from the whole sample thickness contribute to the low-frequency image ($\left|\bm{q}\right|=q_1$, panel e)).

One can quantitatively account for this effect through the kernel $K_z(\bm{r})$ weighting different $\Sigma_z$ planes in Eq.~\ref{eqn:intensity}.
Taking the 2D Fourier transform of Eq.~\ref{eqn:intensity}, under the assumptions of a uniform illuminating field one obtains:
\begin{equation}
    \tilde{I}(\bm{q}) \propto \delta_D(\bm{q}) + \int dz \tilde{K}_z(\bm{q}) \tilde{n}(\bm{q}, z) \,,
    \label{eqn:intensity_fourier}
\end{equation}
where $\delta_D(\bm{q})$ is again Dirac's delta, and $\tilde{K}_z(\bm{q})$ and $\tilde{n}(\bm{q}, z)$ are the 2D Fourier transform with respect to $\bm{x}$ of $K_z(\bm{x})$ and $n(\bm{x}, z)$, respectively. 

Under shear, a rigid translation introduces a phase term $e^{i q_x \gamma (z-z_s)}$ in the $\tilde{n}(\bm{q}, z)$ term of Eq. \ref{eqn:intensity_fourier}. Thus, starting from Eq. \ref{eqn:intensity_fourier}, and assuming that both $\tilde{K}_z(\bm{q})$ and $\left|\tilde{n}(\bm{q}, z)\right|$ are $z$-independent, one retrieves Eq.~\ref{eqn:DFMshear_naiveCorr} when calculating the degree of correlation introduced in Eq.~\ref{eqn:c_DDM}. By injecting the same expression in the more practical far-field version of the degree of correlation, Eq.~\ref{eqn:c_FF-DDM}, one finds the following expression, which
corresponds to Eq. 10 of Ref.~\cite{comp}:
\begin{equation}
    g_{2,FF-DDM}(\bm{q}, \gamma)-1=\sinc^2\left(q_x \gamma \frac{e}{2}\right)\,.
    \label{eqn:farfield_affine}
\end{equation}
These results describe the case where the depth of focus is much larger than the cell gap. By contrast, the more general case where the depth of focus is comparable to or smaller than the gap can be still described
starting from Eq.~\ref{eqn:intensity_fourier}, but it requires to assume an expression for $\tilde{K}_z(\bm{q})$ which decays with growing $z$.

It is therefore important to gain insight on the actual behavior of $\tilde{K}_z(\bm{q})$ before analyzing the experimental results for the DDM signal coming from an affine deformation.
We propose here an experimental method to determine it. Let's consider a sample containing immobile scatterers: as the position of the object plane is scanned through the sample, the microscope image will change, since the effective scattering volume is continuously displaced. Correlation functions calculated during such a scan will decay when the object plane has been displaced by an amount comparable to the size of the scattering volume, i.e. comparable to the depth of focus $L_f(\bm{q})$. Note that this effect is the analogous of the decay modeled by Eq.~\ref{eqn:DFMtraslCorr_pro}: in both cases, the loss of correlation is due to the renewal of the set of particles contributing to the image, here due to a translation along the optical axis $\hat{u}_z$, while for Eq.~\ref{eqn:DFMtraslCorr_pro} it resulted from a lateral shift along $\hat{u}_x$.  The r.h.s. of Eq.~\ref{eqn:DFMtraslCorr_pro}
is the (squared) autocorrelation function of the kernel $\tilde{K}$ which, for a 2D sample, is simply a rect function, i.e. unity within the field of view and zero outside the imaged region.
Thus, based on the analogy between a lateral shift and a vertical one, one recognizes that the correlation decay due to a vertical translation $\delta_z$ corresponds to the spatial autocorrelation of $\tilde{K}_z(\bm{q})$.

\begin{figure}[htbp]
\centering
  \includegraphics[width=\columnwidth,clip]{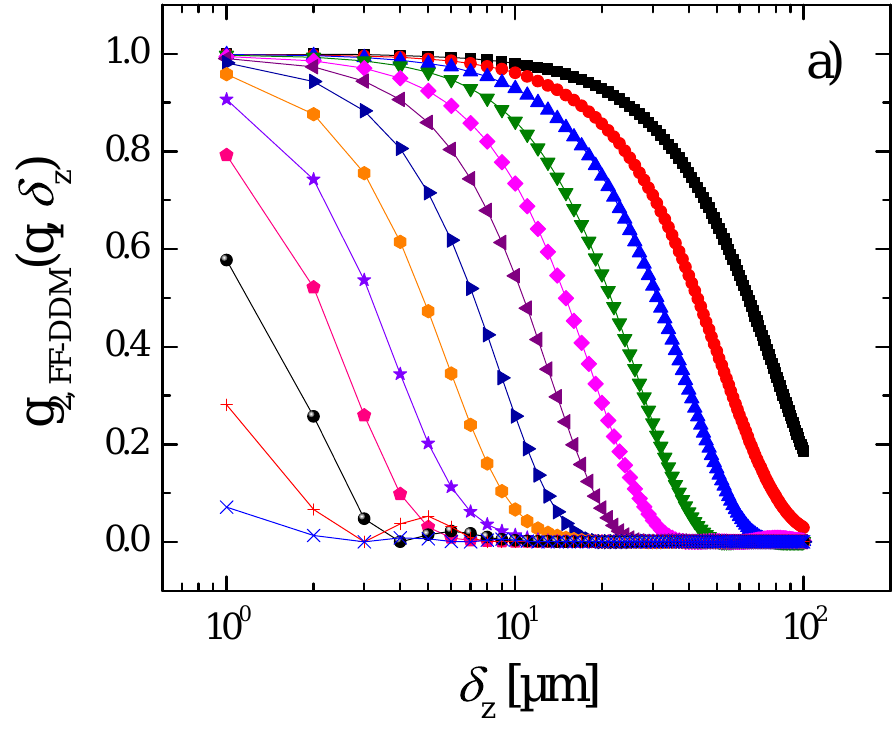}
  \includegraphics[width=\columnwidth,clip]{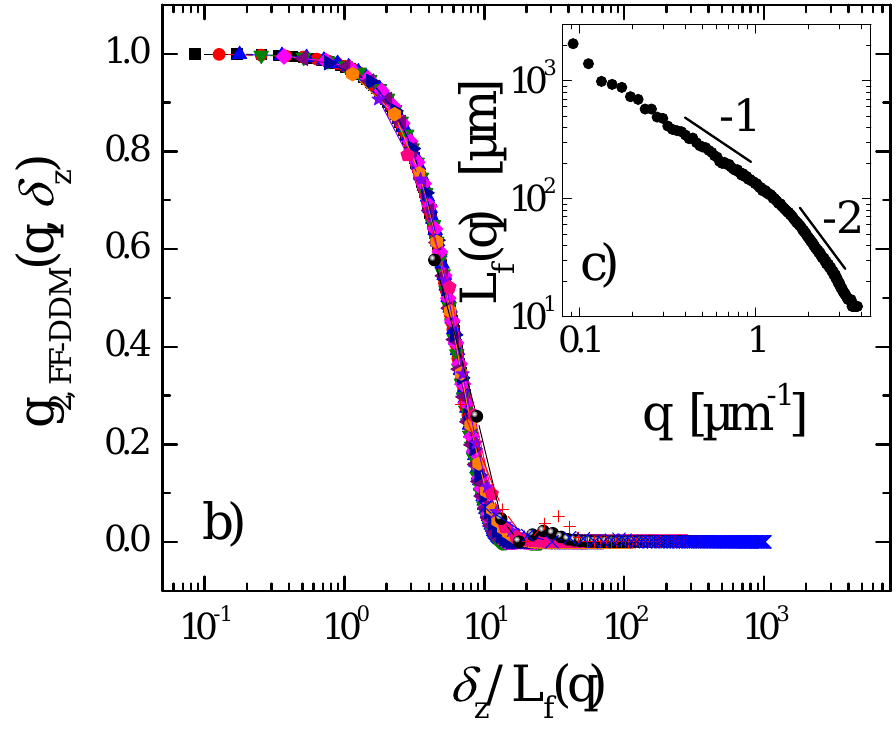}
  \caption{a) Far-field DDM correlation functions between images taken for a sample with frozen dynamics while shifting the object plane by $\delta_z$. From right to left, $q$ increases from $0.1 \um^{-1}$ to $3.8 \um^{-1}$. b):
  master curve obtained from the data shown in a) by normalizing $\delta_z$ with respect to the depth of focus $L_f$. c) $q$ dependence of the depth of focus issued from the scaling procedure:
  two distinct regimes are clearly visible, as discussed in the text.}
  \label{fig:static_effThick}
\end{figure}

Figure~\ref{fig:static_effThick}a shows far-field correlation functions measured for a particle-loaded PA gel, a sample with no measurable spontaneous dynamics, while scanning the sample along $\hat{u}_z$. The correlation functions are plotted versus the shift $\delta_z$ along $\hat{u}_z$, for various $q$ vectors. Note that the decay of $g_{2,FF-CCD}-1$ is faster at larger $q$ vectors, in agreement with the expectation that the effective scattering volume is smaller at higher $q$, as discussed above. To quantify the $q$ dependent depth of focus, we scale data at various $q$ on top of a master curve, by normalizing $\delta_z$ with respect to a scaling factor, which is identified with $L_f$ (Fig.~\ref{fig:static_effThick}b). This procedure determines $L_f(q)$ to within a multiplicative constant that we shall discuss later: for the time being, we note that the choice of ther multiplicative factor used in Fig.~\ref{fig:static_effThick}b provides the correct order of
magnitude for $L_f$, since the correlation functions at all $q$'s decay significantly for a displacement $\delta_z \sim L_f(q)$.
We show in Fig.~\ref{fig:static_effThick}c the $q$ dependence of $L_f$: we find that $L_f$ decreases with $q$, as expected. More in detail, two distinct regimes can be seen: at small $q$, $L_f\sim q^{-1}$, while at large $q$, $L_f\sim q^{-2}$. Giavazzi and coworkers~\cite{Giavazzi2009} have shown that these two regimes stem from two different contributions to the finite depth of focus. For large $q$ vectors, $L_f$ is dictated by the wide spectral range of the white light, whereas at small $q$ vectors the prevailing contribution is due to the finite numerical aperture of the illuminating beam. In most commercial microscopes, this parameter can be tuned by playing with the condenser diaphragm, which allows one to easily change $L_f$.

The procedure illustrated in Fig.~\ref{fig:static_effThick} allows one to determine the sample region $\chi_s(\bm{x'})$ that  effectively contributes to the intensity $I(\bm{x'})$ of a typical bright field microscopy image.
$\chi_s(\bm{x'})$ is obviously centered in $\bm{x}$, the point on the object plane conjugated with $\bm{x'}$. However, its exact shape and size is the nontrivial outcome of the interplay between the finite sample size, the numerical aperture of the objective and the temporal and spatial coherence of the illuminating light.
$\chi_s(\bm{x'})$ can be defined as the set of points $(\bm{x}, z)$ for which the kernel $K_z(\bm{x})$ is significantly different from 0 and it represents the analogous of the scattering volume in a microscopy experiment: the measure of its volume is defined by the integral in the r.h.s. term of Eq. \ref{eqn:intensity}, whereas its thickness along the optical axis sets the effective thickness of the sample probed, which is a key parameter in the experiments under shear.
The shape of $\chi_s(\bm{x'})$ can be sketched starting from the $L_f(\bm{q})$ profile of Fig.~\ref{fig:static_effThick}c: the result is illustrated in Fig.~\ref{fig:focal_depth}h, where the shadowed region representing $\chi_s(\bm{x'})$ corresponds to the intersection between the sample volume and the integral of all conical surfaces for scattering angles $0\leq\theta\leq \pi/2$.

\begin{figure}[htbp]
\centering
  \includegraphics[width=\columnwidth,clip]{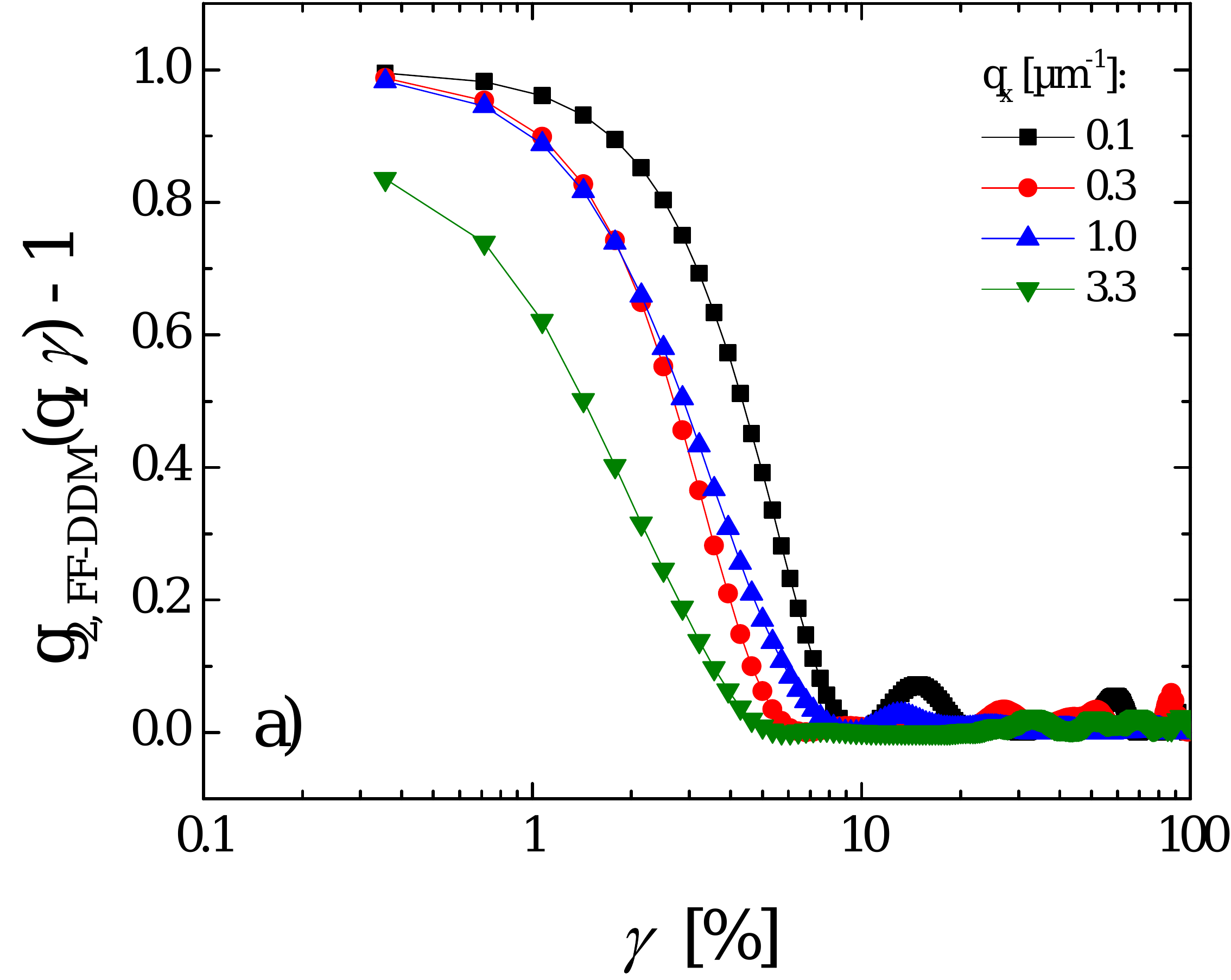}
  \includegraphics[width=\columnwidth,clip]{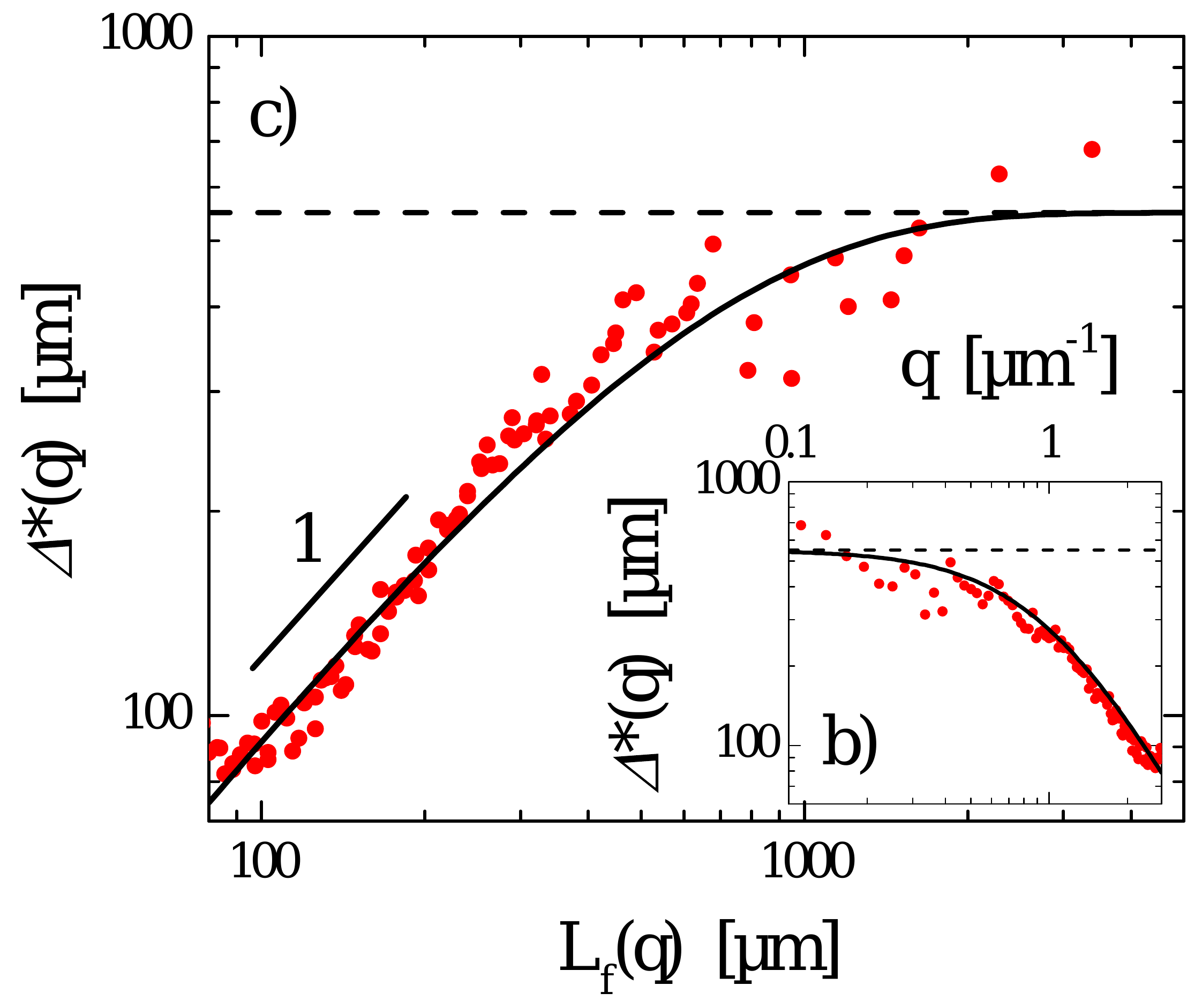}
  \caption{a) Far field DDM correlation functions for a PA gel under shear, as a function of applied strain $\gamma$. Different curves correspond to a few representative scattering vectors $q_x$, as indicated by the label, oriented along the strain direction. b) effective sample thickness $\Delta^*(q)$ for DDM under shear, obtained for each scattering vector $q$ by fitting the correlation functions of Fig.~\ref{fig:affine_corr}, as described in text. The dashed line represents the gap $e$. c) $\Delta^*$ as a function of the depth of focus $L_f$ introduced in Fig.~\ref{fig:static_effThick}b. The solid line is a guide to the eyes: it reduces to the identity $\Delta^* = L_f$ for small values of $L_f$ and it saturates to the gap value $e$ indicated by the dashed line.}
  \label{fig:affine_corr}
\end{figure}

Figure~\ref{fig:static_effThick}c shows that over the range of accessible $q$ vectors $L_f$ varies by more than 2 orders of magnitudes, from $10\um$ to more than 1 mm. Typical sample thicknesses lay in between these two extreme values. Based on Fig.~\ref{fig:static_effThick}, we expect that, due to partial coherence, the decorrelation induced by an affine deformation will no longer depend on the whole sample thickness $e$, but rather on a $q$-dependent effective thickness $\Delta^*(q)$. At the smallest $q$ vectors, where $L_f$ is likely to exceed $e$, we expect that $\Delta^*(q) \sim e$: in this case, we should retrieve Eq. \ref{eqn:farfield_affine}. No guarantee is given a priori that the same functional form is able to describe also the largest scattering vectors, since the precise shape of $g_{2, FF-DDM}$ depends on the whole $\tilde{K}_z(\bm{q})$ distribution, of which $L_f(\bm{q})$ represents just the first moment.
Correlation functions for different scattering vectors $q_x$ along the shear direction are represented in Fig. \ref{fig:affine_corr}.
As expected, experimental data for $q_x \lesssim 0.3\um^{-1}$ are nicely described by Eq.~\ref{eqn:farfield_affine}. A comparison with Fig. \ref{fig:focal_depth} shows that this regime corresponds to the $q$ range where $L_f(q)>e$. By contrast, the correlation functions for $q_x>0.3\um^{-1}$ decay more slowly than expected from the scaling predicted by Eq.~\ref{eqn:farfield_affine}, in qualitative agreement with a more gradual decay of $\tilde{K}_z(\bm{q})$ with increasing $z$.

To account for this slower decay, it is in principle possible to perform a more refined analysis, yielding the full $\tilde{K}_z(\bm{q})$ profile by deconvolving the correlation functions shown in Fig. \ref{fig:static_effThick}a. However, this procedure is delicate and would require very clean experimental data, since data deconvolution is notoriously sensitive to noise. For this reason, in this paper we take a simpler approach, again focusing on the first moment of the correlation functions, i.e. their integral average $\bar{\gamma}(q)$. In this approximation, the effective thickness $\Delta^*(q)$ is defined by $\Delta^*(q)=\bar{\gamma}/q\pi$, where the $\pi$ prefactor is chosen such that $\Delta^*(q)=e$ in the low $q$ limit.
The $q$ dependence of $\Delta^*$ thus obtained is shown in Fig.~\ref{fig:affine_corr}b: at large $q$, the effective thickness decreases with $q$, the same trend already observed for $L_f$. At small $q$, by contrast, the effective thickness tends to saturate at a value close to the sample thickness $e$. Fig.~\ref{fig:affine_corr}c shows $\Delta^*$ as a function of $L_f$: initially, the effective thickness coincides with the depth of focus, but eventually is limited to the sample thickness $e$, when $L_f$ exceeds the latter. As mentioned when discussing Fig.~\ref{fig:static_effThick}, $L_f$ was defined to within a multiplicative constant: in fact, we have chosen this constant such that $L_f = \Delta^*$ in the $L_f <e$ regime.

To summarize this section, we have shown that the partial coherence typical of a microscopy experiment is associated to a finite depth of focus, playing the role of a $q$-dependent effective sample thickness $\Delta^*(q)$. This has a major impact on the observed affine dynamics, which are strongly suppressed as $\Delta^*(q)$ is reduced. The possibility to easily tune $\Delta^*(q)$ by varying the source coherence thus emerges as a unique advantage of DDM under shear. As a drawback of this otherwise appealing feature, an accurate calibration of $\Delta^*(q)$ is crucial to correctly model the affine contribution to the correlation functions. One way to do this is to perform an experiment under shear and to analyze the signal scattered in the direction parallel to the shear, assuming that the affine contribution dominates.
However, such method can be unpractical, since for most samples of physical interest nonaffine contributions may become relevant even at small strains. For this reason, we proposed a convenient experimental protocol to access the depth of focus $L_f(q)$, furthermore showing that $\Delta^*(q)$ essentially coincides with $L_f(q)$ when the latter is smaller than the sample thickness $e$.

\subsection{Nonaffine deformations}

Using the $\Delta^*(q)$ values extracted from calibration data shown in Fig. \ref{fig:affine_corr}, the investigation of nonaffine deformations follows steps similar to those discussed in Ref.~\cite{comp}. We write the total displacement field $\Delta \bm{r}(\bm{r}, \gamma)$ as the sum of an affine term $\Delta \bm{r}^{(aff)}(\bm{r}, \gamma)$ and a nonaffine one $\bm{R'}$, and assume, following previous works~\cite{didonna2005,basu2011} corroborated by particle tracking measurements on our sample~\cite{comp}, that $\bm{R'}$ is isotropic, Gaussian-distributed and uncorrelated from $\Delta \bm{r}^{(aff)}(\bm{r}, \gamma)$. These assumptions allow one to factorize the correlation function in the product of an affine and a nonaffine term.

In Ref. \cite{comp} we reported experimental evidence that DLS correlation functions are well described by the analytic expression
\begin{equation}
    g_{2,DLS}(\bm{q}, \gamma) - 1 =\sinc^2\left(q_x \gamma \frac{e}{2}\right) \exp \left(-\frac{1}{3}q^2 c \gamma^2\right )\,,
    \label{eqn:DLS_nonAffineCorr}
\end{equation}
where $c$ is a generalized diffusion coefficient, defined as the ratio of the mean square nonaffine displacement over the square of the strain.
A straightforward extension of Eq.~\ref{eqn:DLS_nonAffineCorr} to DDM would consist in replacing the sample thickness $e$ by the effective thickness $\Delta^*(q)$. However, as discussed in the previous section, the affine term is no longer described by a $\sinc^2$ function for $\Delta^*<e$. For this reason, we take a more general approach and assume the following functional form for the correlation function:
\begin{equation}
    g_{2,FF-DDM}(\bm{q}, \gamma) - 1 =A\left(q_x \gamma \frac{\Delta^*}{2}\right) \exp \left(-\frac{1}{3}q^2 c \gamma^2\right )\,,
    \label{eqn:DDM_nonAffineCorr}
\end{equation}
where $A(u)$ represents a generic function describing the affine contribution to the loss of correlation, which decays from 1 to 0 as its non-dimensional variable $u$ grows beyond $O(1)$.
Nonaffine deformations can be detected for large enough $q$ values (otherwise the correlation function is insensitive to nonaffine displacements), provided that the $q_x$ component parallel to the shear direction is small enough (otherwise affine motion dominates the decay of the correlator). More precisely, the analysis should be performed on regions in $q$ space centered around the $q_x=0$ axis, with a width $q_w$ satisfying $q_w\Delta^*/q < \sqrt{c}$.

\begin{figure}[htbp]
\centering
  \includegraphics[width=\columnwidth,clip]{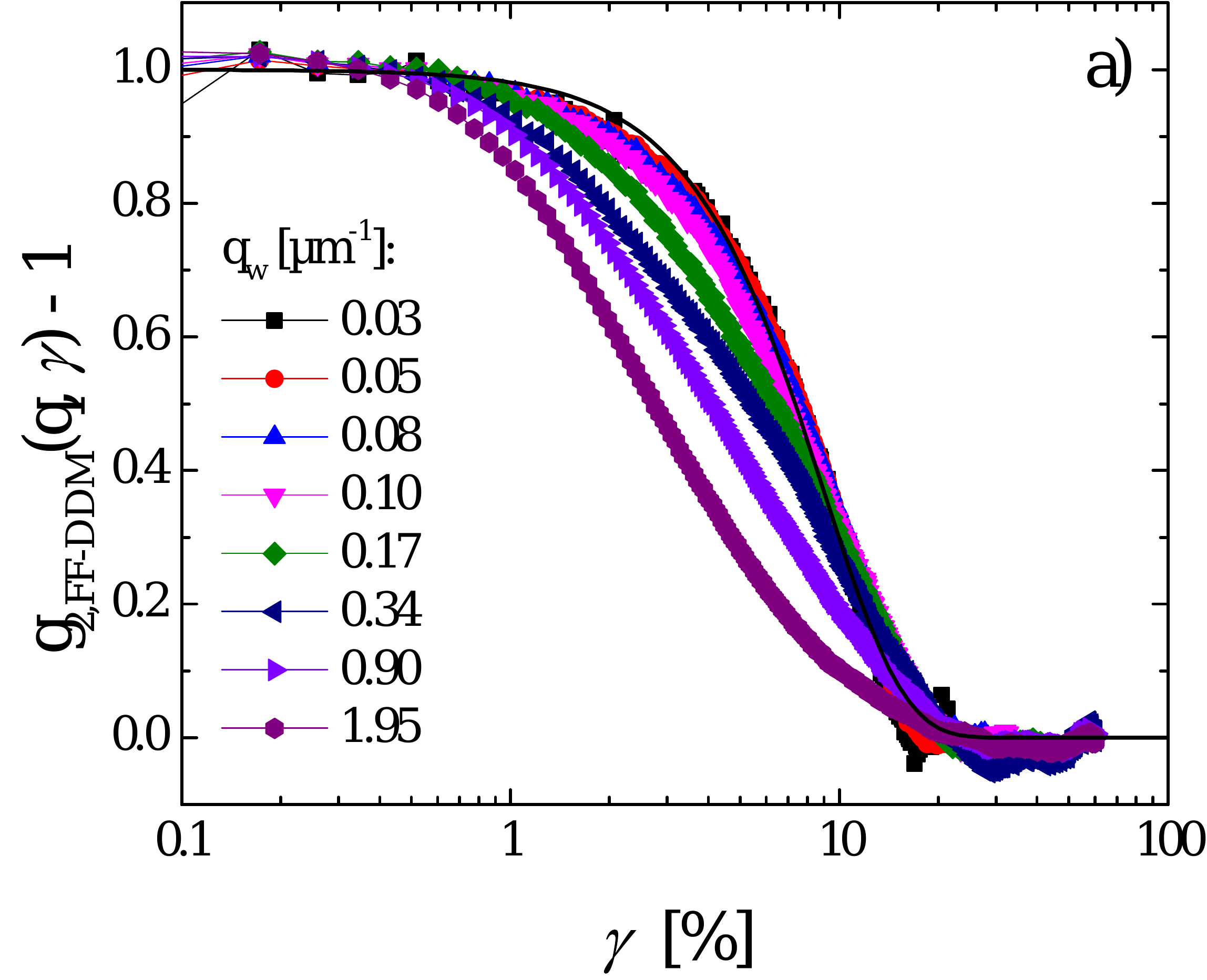}
  \includegraphics[width=\columnwidth,clip]{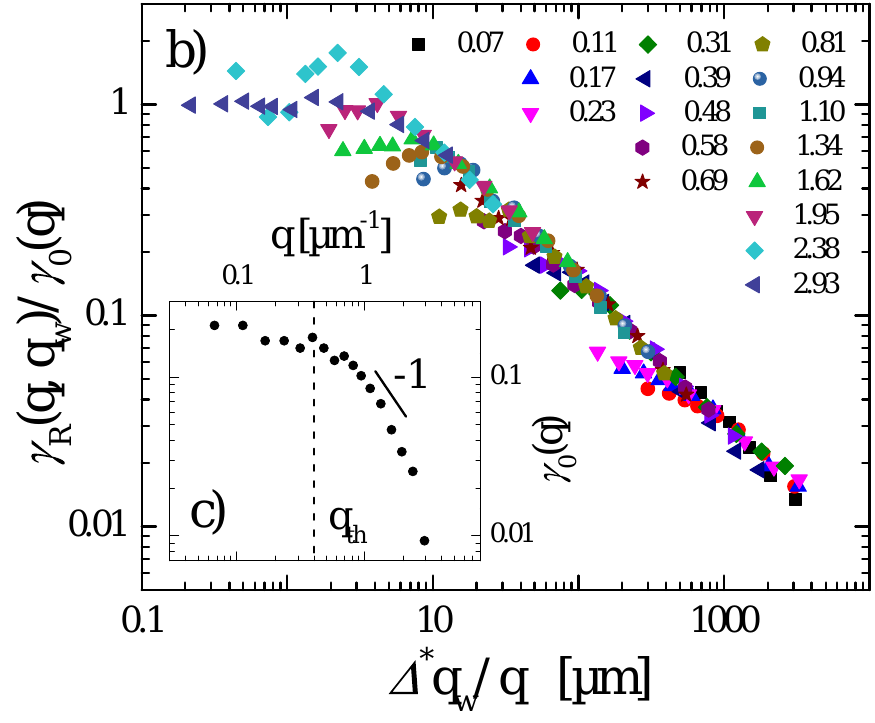}
  \caption{a) Representative far field correlation functions for $q=1.95\um^{-1}$. The scattering vector is mainly oriented in the $\hat{u}_y$ direction, but has a small component along the shear direction, due to the finite width $q_w$ of region in Fourier space used to compute $g_{2,FF-DDM}-1$. Symbols refer to data for the same $q$ but different $q_w$, as indicated by the labels. The line is the Gaussian decay predicted by Eq.~\ref{eqn:DDM_nonAffineCorr}, with a nonaffine parameter $c\approx 8\um^2$.
  b) Rescaled relaxation strains plotted against the scaling variable $\Delta^*q_x/q$.
Note the tendency for data at small scattering vectors to deviate from the scaling law (see e.g. the pink down triangles for $\Delta^*q_x/q \lesssim 200 \um$), which is due to finite scattering volume effects. c) Vertical shift factors $\gamma_0(q)$ used to obtain the master curve shown in b). Two regimes are observed, as detailed in the text.}
  \label{fig:DDM_nonaffRescale}
\end{figure}

To test this approach, the dynamics under shear of the same polyacrylamide gel of Ref.~\cite{comp} are measured with DDM, calculating the far field correlation functions $g_{2,FF-DDM}(\bm{q}, \gamma) - 1$. Representative curves are shown in Fig.~\ref{fig:DDM_nonaffRescale}a for $q=1.95\um^{-1}$ mainly oriented in the $\hat{u}_y$ direction but with a small $q_w$ component (indicated by the labels), due to the finite size of the regions in Fourier space over which $g_{2,FF-DDM}-1$ is averaged. In the $q_w\rightarrow 0$ limit (practically reached for $q_w \lesssim 0.1\um^{-1}$), the correlation decay is well described by a Gaussian, in agreement with Eq. \ref{eqn:DDM_nonAffineCorr}, whereas for large $q_w$ the contribution of affine displacement becomes increasingly relevant.

As already suggested in Ref.~\cite{comp}, it is instructive to rescale, for a given choice of $q_w$, the relaxation strain $\gamma_R(q, q_w)$ (here calculated as the integral average of $g_{2,FF-DDM}(\bm{q}, \gamma) - 1$) by the relaxation strain $\gamma_0(q)$ in the $q_w \rightarrow 0$ limit. The scaled relaxation strain is plotted in Fig.~\ref{fig:DDM_nonaffRescale}b against the scaling variable $\Delta^* q_w/q$. Similarly to Fig. 7 of Ref.~\cite{comp}, the scaled relaxation strain shows a decreasing, affinity-dominated trend for large $\Delta^* q_w/q$ and saturates to a plateau for $q_w\Delta^*/q < \sqrt{c}$. The (broad) crossover between the two regimes is located around $q_w\Delta^*/q=4\um$, from which a value $c\approx 16\um^2$ can be extracted, of the same order of magnitude of $c \approx 8\um^2$ obtained from the data at one single scattering vector shown in Fig.~\ref{fig:DDM_nonaffRescale}a. Figure~\ref{fig:DDM_nonaffRescale}c shows the vertical shift factors $\gamma_0$ used to obtain the master curve shown in b), as a function of $q$. At large $q$, the $q^{-1}$ power law expected from Eq.~\ref{eqn:DDM_nonAffineCorr} is found, whereas a plateau is observed at smaller $q$. The small $q$ regime is due to the effect of the finite size $L$ of the field of view, which was not included in Eq.~\ref{eqn:DDM_nonAffineCorr} for simplicity. Finite size effects introduce an additional factor of $\left( 1 - |\delta|/L\right)^2$ in the decay of $g_{2,FF-DDM}-1$, see Eq.~\ref{eqn:DFMtraslCorr_pro}. By comparing Eqs.~\ref{eqn:DFMtraslCorr_pro} and~\ref{eqn:DDM_nonAffineCorr}, one finds that the crossover $q_{th}$ separating the two regimes satisfies $\Delta^*(q_{th})/q_{th}=L\sqrt{c}$. Using $c =8 \um^2$ as obtained from Fig.~\ref{fig:DDM_nonaffRescale}a) one finds $q_{th} = 0.4\um^{-1}$ (dashed line in Fig.~\ref{fig:DDM_nonaffRescale}c), in excellent agreement with the data.

It is worth commenting the value of the generalized diffusion coefficient $c\approx 8-16\um^2$ obtained from Figs.~\ref{fig:DDM_nonaffRescale}a,b. This value is somehow smaller than $c \approx 20-30\um^2$ as measured by DLS for the same system~\cite{comp}, albeit on a different sample. Although relatively large error bars and significant sample to sample fluctuations might explain this discrepancy, we cannot rule out that the difference also stems from the specific features of the two techniques. In particular, because of the finite focal depth of the microscope, DDM is much less sensitive to the dynamics close to the plates, which may differ from those in the bulk, and is more sensitive to strain heterogeneities, since the probed volume is smaller. Moreover, DDM cannot probe particle displacements along the optical axis larger than the focal depth. A systematic investigation of the origin of this discrepancy, which is beyond the scope of the present work, could be performed by varying the coherence of the illuminating source in the DDM experiment, thereby varying the depth of focus, bringing the experimental conditions closer or further to those of a DLS experiment.

To summarize this section, we have shown that DDM under shear contains a priori contributions from both affine and non-affine displacements. The former are however negligible if the inequality $q_w\Delta^*/q < \sqrt{c}$ is satisfied, where $q_w$ is the the thickness along the shear direction of the Fourier space region over which the correlation function is averaged, $\Delta^*$ is the effective sample thickness to be determined as explained in Sec.~\ref{affine_deformation}, and $c$ is a generalized diffusion coefficient that quantifies non-affine displacements.

\section{Conclusions}
\label{sec:conclusions}

We have discussed how microscopy images taken under shear may be analyzed in Fourier space in order to investigate shear-induced dynamics, with an emphasis on decoupling the contribution of non-affine rearrangements from that of the affine deformation. Similarly to the results presented in the companion paper~\cite{comp} devoted to DLS, we find that correlation functions measured for a $q$ vector oriented parallel to the direction of the applied shear are sensitive to both affine and non-affine displacements, with the affine displacements typically dominating, whereas non-affine displacements may be isolated by measuring correlation functions for a $q$ vector oriented perpendicular to the shear direction.

The analogy between DDM and DLS comes with no surprise, since both methods probe the dynamics in Fourier space. However, there are several important distinctions between the implementation of both experiments, which must be fully addressed in order to correctly extract the desired information. While the usual DLS correlation functions are well suited for characterizing the shear-induced dynamics~\cite{comp}, the `standard' DDM correlator (here referred to as $c_{DDM}$) is unfit for experiments under shear, because it exhibits wide oscillations due to the rigid displacement of sample planes. This difficulty can be addressed using the `far-field' DDM correlator ($c_{FF-DDM}$) that has the desirable property of being much less sensitive to rigid translations. Another key difference between DLS and DDM is the coherence of the illuminating light. In most microscopy experiments, the source has a limited coherence, which results in a finite and $q$-dependent depth of focus $L_f(q)$. Knowledge of $L_f(q)$ is important, since it strongly affects the relative weight of the affine contribution to the correlation functions. To this end, we have proposed a calibration process based on a vertical scan of the sample. This method allows one to determine the effective sample thickness over which the dynamics are probed.

The calibration of the finite depth of focus represents probably the most delicate and cumbersome step of a DDM experiment under shear. On the other hand, the ability of tuning $L_f$ represents a great opportunity. Tuning the effective thickness of the probed sample allows one to optimize the experimental geometry in order to attain the best sensitivity to nonaffine motion. Moreover, different sample slices across the gap may be scanned, which is crucial to check for strain inhomogeneities, as in shear banding, and to address the role of sample boundaries and the possible occurrence of wall slip. Finally, DDM under shear benefits from the same appealing features that are, in general, making this method an increasingly popular one: the simplicity and widespread availability of the setup (a commercial microscope), as well as the advantages associated with the imaging geometry, allowing one to specialize the analysis to just a portion of sample or to combine the Fourier space analysis to real-space methods such as particle tracking.

\section{Acknowledgments}
This work was funded by Agence National de la Recherche (ANR) Grant ANR-14-CE32-0005-01, Centre National d'\'{e}tudes Spatiales (CNES), and the EU (Marie Sklodowska-Curie ITN Supolen Grant
607937). We thank R. Cerbino, F. Giavazzi and M. Alaimo for several illuminating discussions.

\section*{Appendix 1: list of symbols}

A list of the symbols used in the main text and in Appenix 2 is given below. The list is organized in four separate sections (setup, electric field and detector signal, shear deformation and particles' displacement, correlation functions); within each section the symbols are in alphabetical order.

Experimental setup and geometry

\begin{tabular}{ l l }
	$e$ 			& sample thickness  \\
	$L$			& image size in the $\hat{u}_x$ direction\\
	$l_p$			& pixel size \\
	$M$			& microscope magnification \\
	$N_x, N_y$		& image size (pixels) \\
	$\bm{q}$		& scattering vector \\
	$q_x, q_y$		& $x, y$ components of the scattering vector \\
	$\bm{q}_x, \bm{q}_y$ & scattering vectors oriented along the $x$ and $y$ \\
				& direction respectively \\
	$q_{min}$		& $=2\pi M /L$, scattering vector resolution in the \\
				& shear direction, set by the image size\\
	$q_w$		& size of the analyzed region (in $q$ space) along $\hat{u}_x$ \\
	$t$			& time \\
	$\hat{u}_x, \hat{u}_y, \hat{u}_z$ & unitary vectors oriented along the shear, vorticity \\
				& and shear gradient directions, respectively. \\
				& $\hat{u}_z$ also coincides with the optical axis direction. \\
	$z$			& position in the sample along the optical axis, \\
				& relative to the object plane position ($z=0$) \\
	$\chi_s$		& scattering volume\\
	$\tau$ 		& time delay between two images\\
\end{tabular}
\newline

Electric field, intensity, detector signal

\begin{tabular}{ l l }
	$B_p$   		&   "static" contribution to $I_p$ coming from \\
				& the optical background \\
	$\tilde{B}, \tilde{S}, \tilde{N}$	& spatial Fourier transforms of $B_p$, $S_p$ and $N_p$.\\
	$\bm{E}(\bm{x}_p)$ & total electric field in position $\bm{x}_p$ \\
	$\bm{E}_{LO}$	& incident electric field (local oscillator)\\
	$\bm{E}_{sc}$	& electric field scattered by the sample \\
	$I_p=I(\bm{x}_p)$ & intensity detected by $p$-th pixel, conjugated \\
				&  with position $\bm{x}_p$ in the sample image plane. \\
	$\tilde{I}(\bm{q})$	& 2D spatial Fourier transform of $I(\bm{x})$ \\
	$\bm{K}_z(\bm{x})$ & point spread function of a particle \\
				& at distance $z$ from the image plane \\
	$\tilde{\bm{K}}_z(\bm{q})$ & 2D spatial Fourier transform of $\bm{K}_z(\bm{x})$ \\
				& with respect to $\bm{x}$ \\
	$N_p$    		&   camera noise contribution to $I_p$ \\
	$S_p$    		&   "dynamic" contribution to $I_p$ coming from \\
				& the light scattered by the sample \\
	$\sigma_N^2$	& time variance of the noise signal $N_p$ \\
\end{tabular}
\newline

Shear deformation, particle displacements

\begin{tabular}{ l l }
	$c$			& generalized diffusion coefficient (non-affine\\
                			& mean squared displacement per unit squared strain)\\
	$L_f$			& depth of focus \\
	$\bm{r}$		& coordinates in the scattering volume \\
	$z_s$			& position of the stagnation plane under shear \\
	$\bm{\delta}$	& translation of a 2D sample along $\hat{u}_x$ \\
	$\delta_z$		& axial shift of the object plane when correlating\\
				& pairs of images to evaluate $L_f$ \\
	$\Delta^*(q)$	& $q$-dependent effective thickness for \\
				& affine decorrelation, Eq.\ref{eqn:DDM_nonAffineCorr} \\
	$\Delta\bm{r}(\bm{r}, \gamma)$ & particle displacement field following \\
				& a shear deformation $\gamma$ \\
	$\Delta\bm{r}^{(aff)}$	& affine displacement field, Eq. \ref{eqn:AffineDeformation} \\
	$\gamma$ 		& macroscopic shear deformation \\
	$\gamma_R(\bm{q}, q_w)$ & deformation needed to decorrelate the signal\\
               			 & in presence of nonaffinities\\
	$\gamma_0(\bm{q})$ & $=\gamma_R(\bm{q}, 0)$ (thin ROI limit) \\
	$\Sigma_z$		& sample plane perpendicular to the optical axis \\
\end{tabular}
\newline

Correlation functions

\begin{tabular}{ l l }
	$c_{DDM}(\bm{q},t,\tau)$ & standard degree of correlation \\
	$c^{corr}_{DDM}(\bm{q},t,\tau)$ & normalized version of $c_{DDM}$, \\
				&decaying from 1 to 0 \\
	$c_{FF-DDM}(\bm{q},t,\tau)$ & far-field degree of correlation \\
	$c^{corr}_{FF-DDM}(\bm{q},t,\tau)$ & normalized version of $c_{FF-DDM}$, \\
				& decaying from 1 to 0 \\
	$g_{2,DDM}(\bm{q},\tau)-1$ & DDM intensity correlation function\\ 
	$g_{2,FF-DDM}(\bm{q},\tau)-1$ & far-field DDM intensity correlation \\
	$\varepsilon_0(q)$	& deviation from 1 of the $\tau\rightarrow 0^+$ \\
				& limit of $c_{DDM}$ \\
	$\varepsilon_\infty(q)$ & $\tau\rightarrow \infty$ limit of $c_{DDM}$ \\

\end{tabular}

\section*{Appendix 2: DDM correlation functions}

Let's $I_p(t)$ be the signal detected by $p$-th pixel of the camera at time $t$. One can distinguish three different contributions to the signal: $I_p(t)=B_p+S_p(t)+N_p(t)$, $B_p$ being a static optical background, $N_p(t)$ the camera noise, and $S_p(t)$ a dynamic signal related to the light scattered by the sample. These three contributions are mutually uncorrelated (e.g. $\left<S_p(t)N_{p'}(t')\right>=\left<S_p(t)\right>\left<N_{p'}(t')\right>$), and both $S_p(t)$ and $N_p(t)$ have zero time average. Moreover, we assume that the camera noise is uncorrelated in space and time, and only characterized by its time variance $\sigma_N^2$: $\left<N_p(t)N_{p'}(t')\right>=\sigma_N^2\delta_{D,p, p'}\delta_D(t-t')$.

Recalling the formalism introduced in Sec. \ref{sec:DDM}, one can identify $B_p$ with $\left|\bm{E}_{LO}(\bm{x}_p)\right|^2$, $\bm{x}_p$ being the position of the point in the imaged plane conjugated with the $p$-th pixel. Moreover, $S_p(t)=2\Re\left[\bm{E}_{LO}(\bm{x}_p)\bm{E}_{sc}^*(\bm{x}_p, t)\right]$, since the first Born approximation, characterizing the single scattering regime, allows the quadratic term in $\bm{E}_{sc}$ to be dropped. By further expressing the scattering field $\bm{E}_{sc}$ in terms of a sum over the contributions of individual particles convoluted with their point spread function $\bm{K}_z(\bm{x})$, one obtains $S_p(t)=2\Re\left[\sum_j\bm{E}_{LO}(\bm{x}_p)\bm{K}_{z_j}^*(\bm{x}_p - \bm{x}_j(t))\right]$, where $j$ runs on all particles in the scattering volume, and $\bm{x}_j$ and $z_j$ represent the $j$-th particle position projected on the imaged plane and along the optical axis, respectively.

DDM is based on a spatial Fourier analysis of the intensity $I_p(t)$. The Fourier transform of the signal is $\tilde{I}(\bm{q}, t) =\sum_p I_p(t)e^{-i\bm{q}\cdot\bm{x}_p}=\tilde{B}(q)+\tilde{S}(q, t)+\tilde{N}(t)$, where $\tilde{B}(q)$, $\tilde{S}(q, t)$ and $\tilde{N}(t)$ represent the spatial Fourier transforms of $B_p$, $S_p(t)$ and $N_p(t)$, respectively.

Using the statistical properties mentioned before one can show that:
\begin{equation}
\begin{split}
    c_{DDM}(\bm{q}, t, \tau) &= \Re\frac{\left<\tilde{I}(\bm{q},t)\tilde{I}^*(\bm{q}, t+\tau)\right>}{\left<\left|\tilde{I}(\bm{q},t)\right|^2\right>} \\
	&= \Re \frac{\left|B(q)\right|^2 + \left<\tilde{S}(\bm{q}, t)\tilde{S}^*(\bm{q}, t+\tau)\right> + \sigma_N^2\delta(\tau)}
    		{\left|B(q)\right|^2 + \left<\left|\tilde{S}(\bm{q},t)\right|^2\right> + \sigma_N^2}\,,
    \label{eqn:cDDM_derived}
\end{split}
\end{equation}
where ergodicity has been assumed for simplicity.
Because the convolution in the expression of $S_p(t)$ becomes a simple product in Fourier space~\cite{GoodmanIntroductionFourierOptics2005}, one recognizes in $\left<\tilde{S}(\bm{q}, t)\tilde{S}^*(\bm{q}, t+\tau)\right>$ a field correlation function (see Ref.\cite{BernePecora}), whose normalization we now discuss.

By definition, Eq.~\ref{eqn:cDDM_derived} is equal to 1 for $\tau=0$, although as a consequence of camera noise one has $\lim_{\tau\rightarrow 0^+} c_{DDM}(\bm{q}, t, \tau) = 1-\varepsilon_0(\bm{q})$, with $\varepsilon_0(\bm{q})=\sigma_N^2 / \left<\left|\tilde{I}(\bm{q},t)\right|^2\right>$. On the other hand, for $\tau$ much larger than the correlation time, the r.h.s. of Eq.~\ref{eqn:cDDM_derived} tends to a plateau whose value corresponds to the ratio between static background and total signal: $\varepsilon_\infty(\bm{q})= \left|B(q)\right|^2 / \left<\left|\tilde{I}(\bm{q},t)\right|^2\right>$. In our experiments, $\varepsilon_0(\bm{q})$ and $\varepsilon_\infty(\bm{q})$ are measured and used to obtain a corrected correlation function: $c^{corr}_{DDM}(\bm{q}, t, \tau) = \left[c_{DDM}(\bm{q}, t, \tau) - \varepsilon_\infty(\bm{q})\right]/\left[1-\varepsilon_0(\bm{q})-\varepsilon_\infty(\bm{q})\right]$.

The same approach allows one to explicitly derive an expression for $c_{FF-DDM}(\bm{q},t,\tau)$, which will involve many more cross terms. Neglecting the camera noise one obtains: $c_{FF-DDM}(\bm{q}, t, \tau) \propto \left<\left|\tilde{S}(\bm{q}, t)\tilde{S}^*(\bm{q}, t+\tau)\right|^2\right>-\left[\left<\left|\tilde{S}(\bm{q},t)\right|^2\right>\right]^2+2\left|B(q)\right|^2\Re\left<\tilde{S}(\bm{q}, t)\tilde{S}^*(\bm{q}, t+\tau)\right>$. The last term turns out to be negligible in the range of scattering vectors $q$ where the signal dominates the static background: this is typically the case except at vanishingly small $q$. In this regime one obtains:

\begin{equation}
    c_{FF-DDM}(\bm{q}, t, \tau) \approx \frac{\left<\left|\tilde{S}(\bm{q}, t)\tilde{S}^*(\bm{q}, t+\tau)\right|^2\right>-\left[\left<\left|\tilde{S}(\bm{q},t)\right|^2\right>\right]^2}
    {\left[\left<\left|\tilde{I}(\bm{q},t)\right|^2\right>\right]^2}
    \label{eqn:cFFDDM_derived}
\end{equation}

Since $\tilde{S}(\bm{q}, t)$ is proportional to the scattered field, Eq. \ref{eqn:cFFDDM_derived} turns out to be an intensity correlation function: it decays from a value slightly lower than 1 for $\tau \rightarrow 0$ (because of the contributions of $\tilde{B}(q)$ and $\tilde{N}(t)$ to the denominator) to 0 for $\tau \rightarrow \infty$. As for Eq.~\ref{eqn:cDDM_derived}, the raw correlation functions can be corrected in order to obtain a correlation function $c^{corr}_{FF-DDM}(\bm{q}, t, \tau)$ decaying from 1 to 0. All data presented in the main text have been corrected in this way; the superscript $corr$ has been omitted to avoid overburdening the notation.

\footnotesize{


\providecommand*{\mcitethebibliography}{\thebibliography}
\csname @ifundefined\endcsname{endmcitethebibliography}
{\let\endmcitethebibliography\endthebibliography}{}

}

\end{document}